# Reviewed of the compression limit of an individual sequence using the Set Shaping Theory

Aida Koch, Alix Petit

Abstract: In this article, we will analyze in detail the coding limit of an individual sequence by introducing the latest developments brought by the Set Shaping Theory. This new theory made us realize that there is a huge difference between source entropy and zero order empirical entropy. Understanding the differences between these two variables allows us to take an important step forward in the study of the compression limit of an individual sequence, which we know is not calculable.

## Introduction

In this article, we will analyze in detail the coding limit of an individual sequence by introducing the latest developments brought by the Set Shaping Theory. This new theory made us realize that there is a huge difference between source entropy and zero order empirical entropy. Let us start by asking the following question: What is the compression limit of an individual sequence?

Two answers can be given to this question:

**1) The compression limit of an individual sequence is an unsolved problem.** Furthermore, it has been shown that this limit cannot be calculated.

**2) Shannon defined a subproblem of great practical relevance, in which it is possible to define a theoretical limit (source coding theorem).** This subproblem is purely probabilistic and is called source coding.

## Definition of the information transfer problem in its most general form

To understand these two answers it is essential to introduce the problem of information transmission in its most general form. For the transmission of information to take place between two points, it is essential that the two subjects share a communication language. This requirement is necessary, without a language of communication the transmitted message is not understood and therefore does not carry any information. **Consequently, the information transmission problem, in its most general form, studies the transfer of information in any form between two points that share a language of communication.**

Contact author: aida.koch445@outlook.com

The problem with this definition is that the concept of "information in any form" is difficult to formalize mathematically. **For this reason it is essential to introduce a first simplification of the problem, in which the information is defined as a numerical sequence.** In this way, a subproblem is obtained, in which the concept of information is well defined and can be formalized from the mathematical point of view.

Consequently, the two elements necessary for the transfer of information are: a compressed sequence which carries the information and a communication language known by both the encoder and the decoder.

**Compressed sequence:** represents any information derived from the sequence to be transmitted. From a practical point of view, any information obtained from the sequence that is useful for the decoder to go back to the original message must be considered part of the compressed message.

**Communication language:** represents a set of rules that allows communication and therefore the transfer of information between two subjects. From a practical point of view, any information present before the generation of the message can be considered part of the communication language.

## Source coding

**Shannon in his famous article [1] introduces a further modification to the information transfer problem, which allows him to obtain a subproblem in which it is possible to define the theoretical compression limit.** For this purpose, Shannon introduces the source that generates the message. Furthermore, the coding scheme which converts the symbols into codewords is generated on the information of the source and not on the information obtained from the message.**Therefore, the coding scheme being defined before the generation of the message can be considered as an integral part of the communication language.** Then, Shannon formalizes the concept of information, for this purpose he develops a variable that uniquely defines the message transmitted in the presence of an encoding scheme. The solution developed by Shannon is to consider information as a measure of a variation. The function generated that performs this measurement is called entropy $H$. The following sentence, taken from the article "A Mathematical Theory of Communication", makes us understand Shannon's idea of entropy.

*"In the limiting case where one probability is unity (certainty) and all the others zero (impossibility), then H is zero (no uncertainty at all- no freedom of choice - no information)."*

**With regard to the subproblem he defined, Shannon manages to define a theoretical compression limit by developing the source coding theorem.** This theorem proves that entropy defines the minimum average length of the codewords that replace the symbols in the

Contact author: aida.koch445@outlook.com

compressed message. Thus, given a source *X* of random variables i.i.d. which generates a message of length *N* the compressed message cannot be longer than *NH(X)*. The encoding scheme being defined before message generation is not part of the compressed message.

## Application of the theory developed by Shannon when the source is not known

When the source is not known, it is possible to calculate the zero order empirical entropy $NH_{0(m)}$[2], in which the frequencies are obtained from the message m to be transmitted. In this case, using the frequencies of the symbols in the sequence the coding scheme is developed after the message has been generated therefore it must be part of the compressed message. Consequently, the compressed message is defined by $NH_{0(m)} + encoding\ scheme$. Being the value of $NH_{0(m)}$ very close to the value of *H(X)* the coding scheme represents a redundancy, which is also called inefficiency of the entropy coding. **In practice, the non-knowledge of the source that generates the message determines the presence of a redundancy in the compressed message.**

Furthermore, the length of the coding scheme depends on the language used, therefore, it is not possible to define it precisely. Consequently, it's not possible to define a limit on the length of the compressed message. Therefore, using the formalism introduced by Shannon, we have shown that the compression limit of a sequence whose source is unknown is not computable.

Difference between the source entropy *H(X)* and the zero order empirical entropy $NH_{0(m)}$.
The value of *NH(X)* represents the compressed message. **Therefore, entropy, in this case, defines message information in a more general sense, not just Shannon information.** Unfortunately, this method does not represent the general case of information transmission, but represents a subproblem in which both the encoder and the decoder know the source that generates the message.

The value of $NH_{0(m)}$ does not represent the compressed message. **Therefore, the zero order empirical entropy represents only the average value of the Shannon information of the single symbol.** Consequently, the zero order empirical entropy has a less general meaning than the entropy of the source. However, in this case, the problem of information transmission is being addressed in a much more general way than in the case of source coding.

To demonstrate what has just been said, the following example is given: let us take a uniform source *X* of random variables i.i.d. which generates a message of length *N*. Of course, this sequence cannot be compressed. Thus, on average the compressed message must have a length greater than or equal to *N*. In the first case, the encoder and the decoder know the source and therefore we can apply the source encoding. The result obtained is that all the messages have length *NH(X)* and since *H(X)=1* (we use the dimension of the alphabet as the base of the entropy) the compressed message has a length N. Therefore, in this case, we reach the

Contact author: aida.koch445@outlook.com

theoretical compression limit. In the second case, both the encoder and the decoder do not know the source. Consequently, the encoder must use the value of the frequencies of the symbols in the message to calculate the entropy. In this way, the zero order empirical entropy $NH_0(m)$ is obtained. The source is uniform but only a small amount of the generated messages have uniform symbol distribution. Therefore, the average value of the zero order empirical entropy $NH_0(m)$ of the messages will be less than $H(X)$. Consequently, if we do not consider the coding scheme as part of the compressed message, we obtain an illogical result, in fact we have $NH_0f(m) < N$. In practice, we managed to compress a random sequence. **The error depends on the fact that, in this second case, we obtained the coding scheme after the message was generated therefore, it must be part of the compressed message.** In conclusion, to obtain a correct result, the compressed string must be defined by $NH_0(m) +$ $encoding\ scheme$ scheme whose length is greater than $N$. Therefore, we have a redundancy due to non-knowledge of the source.

## Modern approach to information theory

One of the most important aspects covered is to understand that source coding is a subproblem with respect to the problem of information transmission in its most general form. **Consequently, when the source is not known, only the zero order empirical entropy can be calculated, a parameter that has a much more limited value than the entropy of the source.** Indeed, the value obtained by multiplying the zero order empirical entropy by the message length $NH_0(m)$ does not represent the compressed message. In this case, the coding scheme must also be part of the compressed message, having been defined after the generation of the message and therefore cannot be included in the communication language. This approach represents an important change of point of view. In fact, it was absolutely normal, even when the source was not known, to consider the compressed message only as $NH_0(m)$ without considering the coding scheme as is done with source coding.

This change of point of view was necessary with the arrival of a new theory that makes it possible to reduce the redundancy that is generated when the source is not known. This theory can only be applied when the alphabet is greater than or equal to 3. **The purpose of this theory is to make the encoded sequence $NH_{0(m)}$ and the coding scheme interact with each other more efficiently.** This new theory, called Set Shaping Theory [3] [4], develops a series of transforms that create a one-to-one relationship between elements of different sets.

Now, I will explain the basis of this theory. If we have to perform entropy coding on a sequence of length $N$ and alphabet $A$, whose source we do not know, the only information we have is that the sequence will be part of one of $|A|^N$ possible sequences. **Thus, the only possible way for a transform, to improve on average the length of the compressed sequence (encoded sequence+coding scheme), using an entropic coding, is to transform the set of all possible sequences into a new set of the same size composed of sequences that on average can be**

Contact author: aida.koch445@outlook.com

**compressed in less space.** In this way, even if we do not know the source that generates the sequence, we know that if we apply this transform, we can, on average, obtain a compressed message of smaller length than the compressed message of the untransformed sequence.

The set having these characteristics is the set of dimension $|A|^N$ composed of sequences of length $N+K$. Increasing the length of the sequence also increases the size of the set that includes all possible sequences, which becomes $|A|^{N+K}$. Therefore, from this set it is possible to select a subset of size $|A|^N$ composed of sequences having the smallest value of the zero order empirical entropy.

In this way, we obtain the following result: given a message m of length $N$ generated by a source $X$ (unknown) of random variables i.i.d., applying the described transform $f(m)$, we obtain on average:

$$NH(X) < Avg(N_t H_0(f(m) + coding\ scheme) < Avg(NH_0(m) + conding\ scheme)$$

With $Nt>N$
**Avg(NtH0(m)+coding scheme)** = mean value of the compressed sequence *m+coding scheme*
**Avg(NtH0(m)+coding scheme)** =mean value of the compressed transformed sequence *f(m)+coding scheme*

**$NH(X)$ is the compression limit that is not known, in fact we have set the condition that the source is not known.** As mentioned at the beginning, when the source is not known, in the current state of knowledge, it is not possible to define a compression limit.

Now, we prove the obtained result experimentally. Let's take a sequence of length 3 and alphabet 3, the possible sequences are 27. If we increase the length to 4 and keep the alphabet at 3, the possible sequences become 81. Of these 81, we select the 27 with the smallest value of $H_0(m)$. In this way, the two sets have the same number of elements. Thus, we can define a one-to-one relationship between the sequences of the two sets. The following table shows the 27 elements of the two sets. In the first column, we have message m, in the second column, we have $NH_0(m)$, in the third column, we have the transformed message *f(m)* and in the fourth column, we have $NtH_0(f(m))$ of the transformed message *f(m)*.

Contact author:  aida.koch445@outlook.com

| m | | | $NH_{0(m)}$ $N = 3$ | f(m) | | | | $N_t H_0(f(m))$ $N_t = 4$ |
|---|---|---|---|---|---|---|---|---|
| 1 | 1 | 1 | 0.000 | 1 | 1 | 1 | 1 | 0.00 |
| 2 | 2 | 2 | 0.000 | 2 | 2 | 2 | 2 | 0.00 |
| 3 | 3 | 3 | 0.000 | 3 | 3 | 3 | 3 | 0.00 |
| 2 | 1 | 1 | 2.755 | 2 | 1 | 1 | 1 | 3.245 |
| 3 | 1 | 1 | 2.755 | 3 | 1 | 1 | 1 | 3.245 |
| 1 | 2 | 1 | 2.755 | 1 | 2 | 1 | 1 | 3.245 |
| 2 | 2 | 1 | 2.755 | 1 | 3 | 1 | 1 | 3.245 |
| 1 | 3 | 1 | 2.755 | 1 | 1 | 2 | 1 | 3.245 |
| 3 | 3 | 1 | 2.755 | 2 | 2 | 2 | 1 | 3.245 |
| 1 | 1 | 2 | 2.755 | 1 | 1 | 3 | 1 | 3.245 |
| 2 | 1 | 2 | 2.755 | 3 | 3 | 3 | 1 | 3.245 |
| 1 | 2 | 2 | 2.755 | 1 | 1 | 1 | 2 | 3.245 |
| 3 | 2 | 2 | 2.755 | 2 | 2 | 1 | 2 | 3.245 |
| 2 | 3 | 2 | 2.755 | 2 | 1 | 2 | 2 | 3.245 |
| 3 | 3 | 2 | 2.755 | 1 | 2 | 2 | 2 | 3.245 |
| 1 | 1 | 3 | 2.755 | 3 | 2 | 2 | 2 | 3.245 |
| 3 | 1 | 3 | 2.755 | 2 | 3 | 2 | 2 | 3.245 |
| 2 | 2 | 3 | 2.755 | 2 | 2 | 3 | 2 | 3.245 |
| 3 | 2 | 3 | 2.755 | 3 | 3 | 3 | 2 | 3.245 |
| 1 | 3 | 3 | 2.755 | 1 | 1 | 1 | 3 | 3.245 |
| 2 | 3 | 3 | 2.755 | 3 | 3 | 1 | 3 | 3.245 |
| 3 | 2 | 1 | 4.755 | 2 | 2 | 2 | 3 | 3.245 |
| 2 | 3 | 1 | 4.755 | 3 | 3 | 2 | 3 | 3.245 |
| 3 | 1 | 2 | 4.755 | 3 | 1 | 3 | 3 | 3.245 |
| 1 | 3 | 2 | 4.755 | 3 | 2 | 3 | 3 | 3.245 |
| 2 | 1 | 3 | 4.755 | 1 | 3 | 3 | 3 | 3.245 |
| 1 | 2 | 3 | 4.755 | 2 | 3 | 3 | 3 | 3.245 |

The average value of zero order empirical entropy multiplied by the string length *N=3* of the messages m is:

**$Avg(NH_0(m))$ =2.893**
with *N=3*.

The average value of zero order empirical entropy multiplied by the length of the string *Nt=4* of the transformed messages f(m) is:

**$Avg(N_t H_0(f(m))$ =2.884**
con *Nt=4*

Thus, you can see that although the transformed messages are longer, the average value of zero order empirical entropy multiplied by the message length is, on average, smaller when the transform is applied. **Consequently, even if we do not know the source, we know that by**

Contact author: aida.koch445@outlook.com

**applying the transform we obtain, on average, a reduction in the value of zero order empirical entropy multiplied by the length of the message.** Therefore, the length of the encoded message (symbols replaced by codewords) decreases.

Now, let's look at the encoding scheme, in this case making an evaluation is more difficult because the length of the encoding scheme depends on the compression method used. But, we know that, from a theoretical point of view, the parameter that most influences the length of the encoding scheme is the size of the alphabet. Therefore, we calculate the average size of the alphabet A in the two sets.

The mean value of the alphabet size |A|, in the case of messages in the first column of the table, is:

*Avg(|A|)=2.1*

The mean value of the alphabet size |A|, in the case of transformed messages *f(m)* in the third column of the table, is:

*Avg(|A|)=1.89*

The result obtained is not valid only for the reported case. **Indeed, whatever the size of the alphabet, the transformed set being composed of longer sequences always has a greater number of type classes with a number of symbols less than |A|.** With type class, we mean a set of strings that all have the same symbol frequency, for example, string 113 and string 131 belong to the same type class in fact, both have symbol frequency 1=2/3 and 3=1/ 4.

Therefore, being on average $NH_0(f(m)) < NH_0(m)$ and coding scheme *f(m) < coding scheme* m, we have experimentally demonstrated the following inequality:

$$NH(X) < Avg(N_t H_0(f(m) + coding\ scheme) < Avg(NH_0(m) + conding\ scheme)$$

In conclusion, the set of transformed messages has an average compressed message length, using entropy encoding, less than the set of the untransformed messages. Consequently, when the source is not known, if the described transform is applied, a reduction of the compressed message is obtained on average. **In this way, we are able to reduce the inefficiency caused by not knowing the source.**

A notable step forward has recently been made regarding this theory, when a group of information theory students managed to develop a transform capable of applying this new technique also for sequences having large length and size of the alphabet [5]. This is an enormous result because, the size of the set that includes all possible sequences given an alphabet *A* and a length *N*, is $|A|^N$, a value that becomes, as the parameters increase, immediately very large, therefore difficult to manage with a calculator. The program is written in Matlab and has been made public and can be downloaded in the Matlab file exchange:

Contact author: aida.koch445@outlook.com

https://www.mathworks.com/matlabcentral/fileexchange/115590-test-sst-huffman-coding?s_tid=FX_rc1_behav

This version of user Kusala Subramani is very interesting because she analyzes in detail the length variation of the Huffman tree (coding scheme), when applying this type of transform. In her analysis she uses the results reported in the article [5].

https://www.mathworks.com/matlabcentral/fileexchange/118615-set_shaping_theory_huffman_analysis_coding_scheme

Using these public programs, you can experimentally validate the reported results by testing the most famous entropy compression algorithms. In fact, one feature of Matlab is that it has a database of functions that perform entropy coding. Thus, it is particularly easy to test the method with different compression methods. The result you will get is an average decrease in the length of the compressed message. In this way, we can reduce redundancy due to non-knowledge of the source.

In conclusion, the question we posed at the beginning, about the existence of a minimum compression length of a sequence, represents a problem that has not yet been solved. However, new developments regarding the theory of information, managing to reduce the inefficiency that is created when we do not know the source, have made it possible to take a significant step forward on this topic.

Contact author: aida.koch445@outlook.com